# Robin: A Web Security Tool

Guilherme Girotto[1] and Avelino F. Zorzo (supervisor)[2]
*Pontifical Catholic University of Rio Grande do Sul*
*School of Technology, Porto Alegre Brazil*
*guilherme.girotto@edu.pucrs.br*[1], *avelino.zorzo@pucrs.br*[2]

***Abstract*— Thanks to the advance of technology, all kinds of applications are becoming more complete and capable of performing complex tasks that save much of our time. But to perform these tasks, applications require that some personal information are shared, for example credit card, bank accounts, email addresses, etc. All these data must be transferred securely between the final user and the institution application. Nonetheless, several applications might contain residual flaws that may be explored by criminals in order to steal users data. Hence, to help information security professionals and developers to perform penetration tests (pentests) on web applications, this paper presents Robin: A Web Security Tool. The tool is also applied to a real case study in which a very dangerous vulnerability was found. This vulnerability is also described in this paper.**

## I. INTRODUCTION

Over the years, software applications have improved the way how companies provide services to their customers. To provide these facilities, software systems are expanding and becoming more complex, and this expansion has the direct consequence of increasing the company *Attack Surface*, *i.e.*, the organizations infrastructure (applications, endpoints, cloud servers, etc.) vs possible attack vectors that each one of these infrastructure subsystems may suffer [1].

Furthermore, hacking practice is also expanding and becoming more popular [2] [3]. The more complex and huge a system is, the more are the chances of finding some vulnerability a developer may have left in the system [4]. Systems are still built by humans, there will certainly have residual faults[1], and these faults are expensive and can also cause several problems [6]. The increase of more *cyber attacks* and specific *tools* focused on exploring breaches [7] - like *botnets*, *Distributed Denial of Service (DDoS)* and even results of crimes (*e.g.*, stolen credit card numbers, compromised hosts) - emphasizes the necessity of new and more complete tools to help developers to find known vulnerabilities during the development process.

In this context, web-based applications are one of the most affected groups, since this kind of applications are the front door of the companies systems, and has a direct relation with the client. *OWASP* Top 10 [8] project, for example, enumerates the 10 most common security breaches that most of the web applications have and their respective impact on the system. Each of these vulnerabilities has its security issues and may lead to serious problems.

To face that problem, in the past, companies would invest most of their resources on manual code-review to remove residual software faults. This strategy has become impracticable on complex systems and is also error-prone since they are executed by humans. Therefore, the security tools market is getting bigger with the increase of systems complexity and also vulnerabilities discoveries. Companies have become aware that investing in software to detect possible breaches on their systems is less expensive than suffering the consequences of getting their systems - or even worse, their data - exposed. But the most popular and complete tools available on the market are proprietary - it means that their source code belongs to a company and cannot be seen, edited and shared by the rest of the community - and generally have expensive licenses.

To help developer and security engineers on finding residual faults left on a web system we introduce Robin: a new open-source web security tool. Open-source software brings a lot of benefits, including cost-effectiveness, fast development, security, and completely free usage. Besides that, open-source code stimulates developers to use and contribute with it, promoting the culture of performing vulnerabilities research and making the available software more secure to everyone.

This paper sections are structured as follows. Section II provides information about the concepts of penetration tests, attack vectors, ethical hacking, and open-source tools. Section III presents some tools and techniques on the context of web application vulnerabilities and the most used and popular tools available on the market. In Section IV we propose our web security tool, presenting its features, development environment and execution Roadmap. Finally, Sections VII and VIII present future development and some conclusions.

[1]We use fault, error, failure as defined in "Basic concepts and taxonomy of dependable and secure computing" [5]

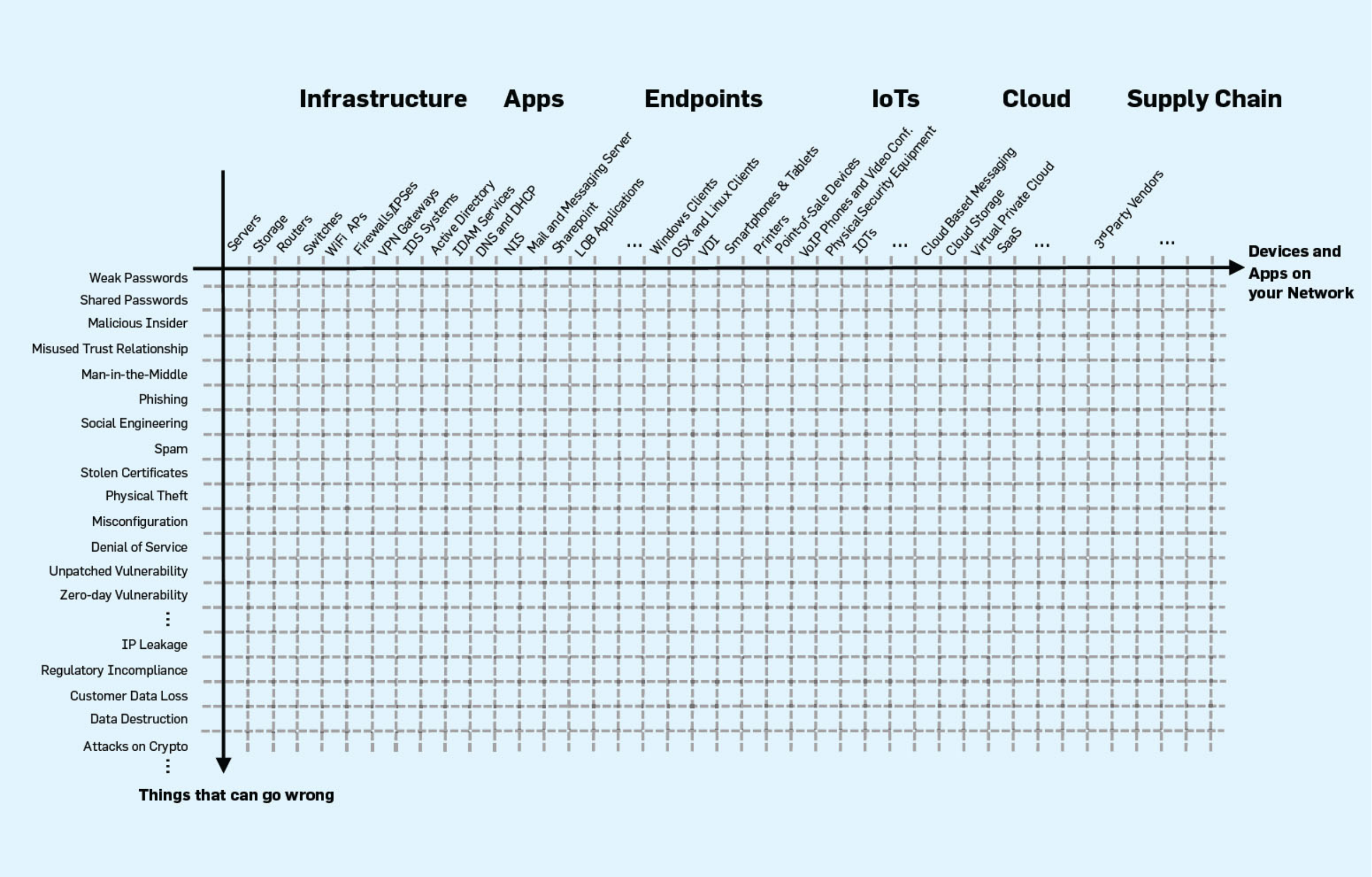

Fig. 1: Enterprise attack surface [1]

## II. BACKGROUND

This section introduces topics that are fundamental to present and discuss the proposal in Section IV. First, Section II-A presents the *Attack Vectors* that affect most companies. After that, in Section II-B, describes *Ethical Hacking* - what it is, why it is important and why it is the central point of our proposal. Finally, Section II-C introduces the *Open Source* topic, presenting some benefits and drawbacks of using and creating open-source projects.

### *A. Attack Surface & Attack Vectors*

As mentioned in Section I, the more a company expands its applications and services, the more *Attack Vectors* may appear due to the company *Attack Surface* increase. A company *Attack Surface* is made up of a relation between its devices and applications connected to the network [1] - summarized as *things can go wrong from a cybersecurity standpoint* - and the *Attack Vectors, i.e, ways in which these things will go wrong*. This relation is illustrated in Figure 1.

### *B. Penetration Testing*

A natural response to the growing cybercrimes is *Penetration Testing (pentesting)* [9] practice. This kind of job focus on performing cyberattacks to detect existing vulnerabilities on a computer-based system. It can be practiced by either well-intended hackers or not. Those that practice *pentesting* to seek to report these vulnerabilities to the software owners so they can fix it are known as *Ethical Hackers* [10]. These well-intentioned hackers focus their efforts on using different *attack vectors* to find vulnerabilities on common software use and report these failures to the software owners so they can fix them before the breach is used to harm the platform users in exchange for bounties. On the companies side, some are adhering to bug bounty programs - such as HackerOne[2] - or even creating their own - as *Facebook Bug Bounty*[3], Apple Security Bounty[4] and others. These programs are summarized on ethical hackers reporting existing vulnerabilities on companies' software and getting paid by their activity.

This new market is a reflection of people using their knowledge to make common software safer. Other amazing initiatives are organizations like OWASP [11], which creates projects like OWASP Top Ten Project [8] to present the most common vulnerabilities mistakes, including examples about how to reproduce, explore and protect against these threats. A knowledge that was used mostly for evilness is now being used to make the web safer, and it must be encouraged.

[2] *https://hackerone.com/*
[3] https://www.facebook.com/whitehat
[4] https://developer.apple.com/security-bounty/

To perform this research, it is very common to use tools and techniques to automate the process of finding and testing possible vulnerabilities. The size of modern systems makes manual analysis and code-review impracticable. Isolated components may become vulnerable when used together, and mapping all possible vulnerable scenarios is unworkable. This impracticability creates a whole new market niche of tools and techniques to improve this field.

### C. Open Source Software

Open Source Software (OSS) is not a new concept. It dates back in the 1950s [12], when hardware and software were delivered together - without a separate fee - and started gaining a lot of relevance with the launch of *GNU Project* by *Richard Stallman* in 1983, who also founded the *Free Software Foundation* a few years later. In 1989, the first version of the *GNU General Public License* was launched, which was used later by popular OSS like *Linux Kernel*.

After years of OSS development, code-sharing, and community engagement, we can see today the benefits - and also the drawbacks - that OSS has. Lorraine *et al.* [13] examined the benefits and drawbacks of OSS usage in 13 companies operating in the secondary software sector in Europe. According to the study, some of the *technical benefits* of using OSS are *Reliability*, *Security*, *Quality*, *Performance* and *Flexibility*. In contrast to the benefits, the *technical drawbacks* of using OSS were *Compatibility Issues*, *Lack of Expertise*, *Poor documentation*, *Proliferation of Interfaces*, *Less Functionality* and *Lack of Roadmaps*. According to the article, the OSS usage has clear benefits and drawbacks. But all drawbacks that were experienced by the companies are characteristics from the OSS that were used by them, and may not be true in other examples of OSS. The *Linux Kernel* is an example of an open-source project that exposes an entire Operational System with a well-documented code-base.

AlMarzouq *et al.* presents [14] the concepts, benefits and challenges of using OSS. On the aspect of the Software-Centered Approach, *i.e.*, the usage of the open-source software, they present similar benefits as discussed above: *Reliability*, *Security* and *Low Cost*. To complement the benefits, they enumerate the challenges that OSS face:

- OSS is not deadline-driven, which may be a problem if the user depends on future events;
- OSS is not well established in some areas as proprietary software is;
- OSS involves in a high entry barrier for non-technical users, since the first cases of success OSS, the users and developers were the same;
- OSS support experience may vary since the OSS maintainers are not obligated to provide support to the users;

## III. RELATED WORK

The creation of tools to analyze web vulnerabilities is not a new trend. Some works discuss and present different tools and techniques to analyze different scenarios. Most of the tools are focused on finding possible attack vectors on web applications, but there are other interesting aspects that focus on developing tools to prevents attacks to the user. Some of the papers that discuss these tools and techniques are presented in *Related Papers*, while the most popular tools available on the market are discussed in *Related Tools*.

### A. Related Papers

Esteban *et al.* [15] present a benchmark of the most common web vulnerability scanners. They argue that these application benchmarks are made periodically, but these analyses are biased by the author's methodology. On this benchmark, they answer questions such as: *The tool truly test and evaluate every vulnerability that it ensures do?*, and even *The tool delivers a real report of all the vulnerabilities tested and exploited?*.

Dalalana and Zorzo [9] provide an overview of *penetration test* scenarios, models, methodologies and tools. They evaluated *1,154* papers with different *pentest* models and tools, classifying them according to their category and phase on the *pentest* execution. Their study goes beyond web-application based tools and approach all categories of *cybersecurity* tools.

Jovanic *et al.* [16] discuss the development of a open-source prototype tool to detect Cross Site Scripting (XSS) vulnerabilities in *PHP* scripts. Their results during the usage of the prototype reported 15 unknown vulnerabilities in three Web applications and reconstruction of 36 known vulnerabilities in three other Web applications, with a false positive of 50%.

Prateek *et al.* [17] propose dynamic analysis techniques to discover *Client Side Vulnerabilities (CSV)*. After that, they create a prototype tool named *FLEX*, which has discovered 11 real-world vulnerabilities.

Sebastian *et al.* [18] present a full automated system capable of detecting DOM-based XSS vulnerabilities in large scale. They applied the system on Alexa top 5000, and the system was able to identify 6,167 vulnerabilities across 418 unique domains, showing that almost 10% of the analyzed websites had some kind of DOM-based Cross Site Scripting (XSS) vulnerabilities. It is interesting how they found a huge amount of vulnerable websites searching only for a specific kind of XSS. The number of vulnerabilities could have been even bigger if they had searched for all kinds of web vulnerabilities.

Philipp *et al.* [19] present a study based on XSS vulnerabilities. They developed a dynamic approach to track the user sensitive data and detect if these data are sent to third party applications. Doing so, they can warn the user about the track of this sensitive information and interrupt the connection if necessary. They also performed some tests in real-word applications and controlled vulnerable scenarios and the tool successfully detected the XSS attacks. A similar tool is presented by Ismail *et al.* [20]. They analyze the user traffic and detect possible XSS vulnerabilities by manipulating these requests.

Tadeusz *et al.* [21] present another way of preventing XSS attacks by manipulating the user string input. They show that most of the injection attacks occur because the user input is manipulated. They perform a context-sensitive string evaluation to prevent undesired modifications on the user input, but it has a drawback of reasonable run-time overhead.

### *B. Related Tools*

Alongside the tools described by the papers in the previous section, we have popular tools that are widely used by most of *infosec* (Information Security) professionals while searching for vulnerabilities. We discuss some of these tools and their main features in the next sections.

*1) Charles:* Charles[5] is a Web Proxy Debug tool. It provides support to create a proxy through the user machine to inspect *HTTP* and *HTTPS* traffic. Charles is not a penetration tool on its own since it does not provide explicit tools and features to perform specific attack scenarios on a penetration test. But it provides an interesting set of functionalities that can help an early phase of detection, such as the content of the packets - even HTTPS packets - and tools to intercept and change requests and responses.

*2) Burp Suite:* Burp Suite[6] is a Cybersecurity Software application with emphasis on providing functionalities to explore pentest scenarios. It has complete and complex tools to perform active scan - search actively on a specific domain for common vulnerabilities -, passive scan, tools to manipulate requests, test for specific attack scenarios and more. This tool drawback is its price, costing $400[7] the individual license and $4000 the enterprise license.

*3) Acunetix:* Acunetix[8] is likely the most popular web vulnerability scanner available on the market. This tool provides an automated scanner that searches for over 6,500 vulnerabilities on software applications with a false positive ration near to 0%. It has a huge set of tools and features that allows performing very custom scan flows to detect the widest range of vulnerabilities on complex systems. Due to its popularity and adoption by the most known and respected software companies, Acunetix has a license price starting at $4,495 for the Standard plan[9] (maximum 5 websites scans).

*4) Arachini:* Arachini[10] has the same purpose as Acunetix - scan websites looking for vulnerabilities. But Arachini is an open-source project, and it is completely free to use. Unfortunately, its development life cycle has officially stopped in January 2020, when the company that mainly maintains the software focus its efforts on investments to hire more developers.

*5) OWASP Zap:* As a reflection of the Top 10 project, the OWASP organization developed *OWASP ZAP* [11]. They provide tools for developers who want to debug their applications and pentesters that are seeking for vulnerabilities on web applications. As the Arachini project, ZAP is an open-source tool and is maintained by the open-source community, with the support of some tech companies, such as the previously mentioned *HackerOne*. Although OWASP ZAP is an open-source tool that acts in the same field as Robin is proposed on. However, we aim to propose a more user friendly tool with a more compact user interface. Robin can also become much more powerful than OWASP ZAP when we consider the 5 Robin modules working together, as described in Section IV.

## IV. PROPOSAL

Given the benefits that Penetration Testing tools bring to Ethical Hackers on automating tasks of finding vulnerabilities on web application systems and the benefits OSS development discussed in Section II, we propose the development of an open-source web penetration tool. We discuss an idealized application - with the most relevant features and techniques used during a penetration test - and an MVP (Minimum Viable Product) implementation, which will be the target of this work.

### *A. Idealized Application*

We visualize the final application as a super-set of the most used and relevant tools during a web application penetration test. Figure 2 illustrates these application modules.

- **Proxy:** A proxy module capable of listing, intercepting and modifying *HTTP* and *HTTPS* requests. This is the fundamental module of the application, and it will be used by all the functionalities that follow.
- **Scanner:** A module that can actively scan a website looking for patterns that are explored by most common vulnerabilities. The module is capable

[5]https://www.charlesproxy.com/
[6]https://portswigger.net/burp
[7]Price in June 2020.
[8]https://www.acunetix.com/
[9]Price in June 2020.
[10]https://www.arachni-scanner.com/
[11]https://owasp.org/www-project-zap/

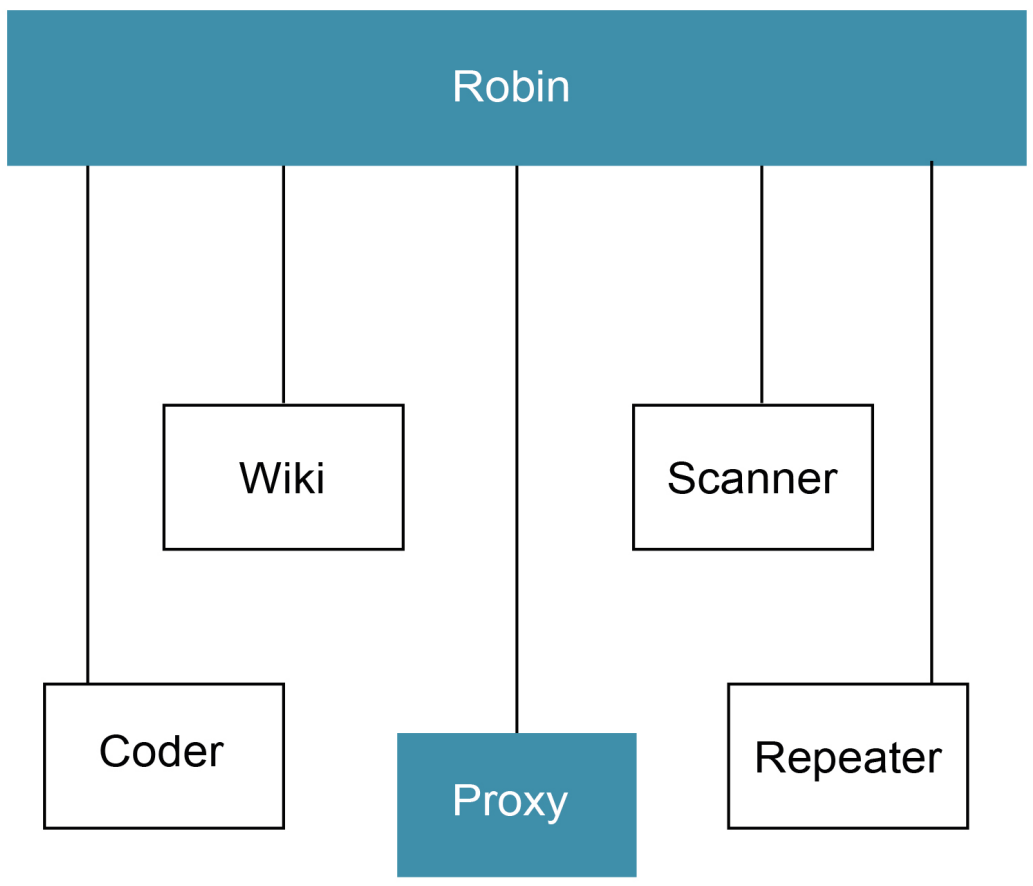

Fig. 2: Application modules

of exposing these patterns and explain a possible attack scenario using the found pattern.

- **Repeater:** A module capable of performing brute-force attacks, following some pre-set configurations and patterns.
- **Coder:** A module capable of encoding and decoding contents using common hashing patterns: *MD5*, *AES*, *SHA256*, *Base 64*, etc. It also could break weak old hash techniques, like reused *One Time Pad*, *MD5*, etc.
- **Wiki:** A module with the explanation of the most common vulnerabilities, how they work, how to explore them and how to protect an application against them, similar to *OWASP Top 10 Project*.

### B. MVP Functionality

In this work, we propose the implementation of an MVP from the idealized application. We will work on the **Proxy** module, highlighted in blue in Figure 2. This implementation consists on the creation of an application that lists, intercepts and edits *HTTP* and *HTTPS* requests. To achieve the functionality, it is necessary to implement the following features:

- Create a local proxy server;
- Redirects all the user network traffic through this proxy;
- Filter traffic to only handle *HTTP* and *HTTPS* requests;
- Create the application certificate - and make the user trust it - to be able to perform a *man in the middle* session and be able to visualize the decrypted *HTTPS* packets content;
- Summarize, format and list these requests;
- Provide a way to intercept and edit the requests before they leave the user machine;

To implement the Proxy module, we use *Electron*[12] - a framework used to build cross-platform applications - and *React JS*[13] - a framework to build user interfaces using *Javascript* and *CSS*. As mentioned on the previous sections, the application is completely free and open-source and is available at Github [14].

## V. DEVELOPMENT

This section presents the development of Robin and its first main module: Proxy. As mentioned in Section IV, the application is built using *Electron* and *React* frameworks. Electron provides a simple way to launch web-based apps as a native OS app. In the same way, React provides a tool to build user interfaces using Javascript. Figure 3 shows a graphical representation of the application core architecture.

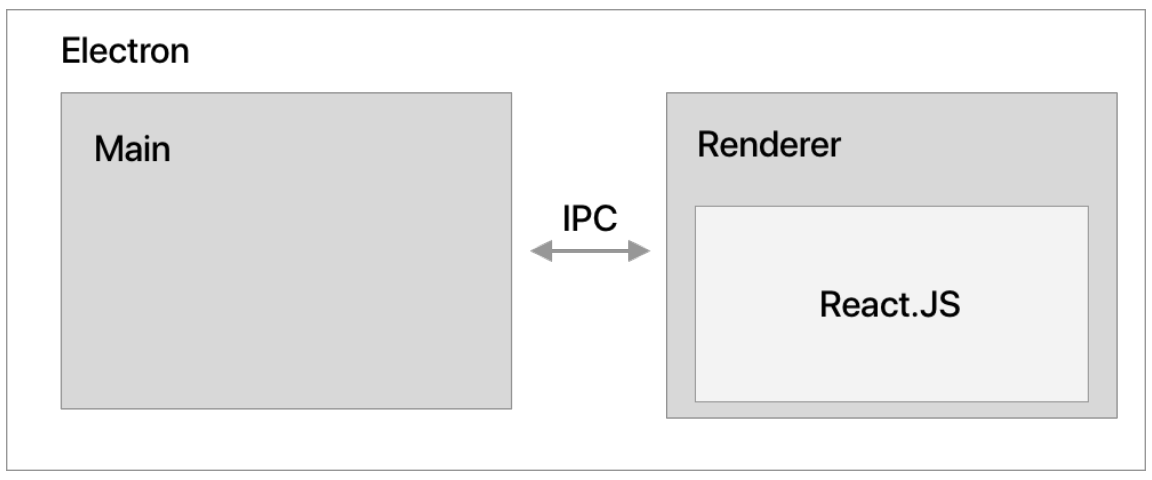

Fig. 3: Application Core Structure

Electron loads two modules: *Main* and *Renderer*. The Main module is the one responsible for performing OS-level tasks, such as intercepting network packages, resolving DNS, turning on machine proxy settings, etc. The Renderer module is responsible for presenting the graphical interface and respond to user interactions. The communication between both modules happens through the Electron IPC (Interprocess Communication) module, which provides a way of making synchronous and asynchronous communication using messages.

As mentioned in Section IV, the main goal of this work is to implement the Proxy module. To create this environment where we can intercept requests before they leave the user machine and the responses before they are consumed by the client, we created *Web Sniffer*[15]: a framework capable of intercepting requests and responses built using Node[16]. The framework architecture is illustrated in Figure 4 and described next:

- The bridge module is responsible for routing the requests to the appropriate servers. If it is an

[12]https://www.electronjs.org/
[13]https://reactjs.org/
[14]https://github.com/
[15]https://github.com/olmps/web-sniffer
[16]https://nodejs.org/

HTTP request, it is redirected to HTTP Server, if it is an HTTPS request, it is redirected to the HTTPS server.

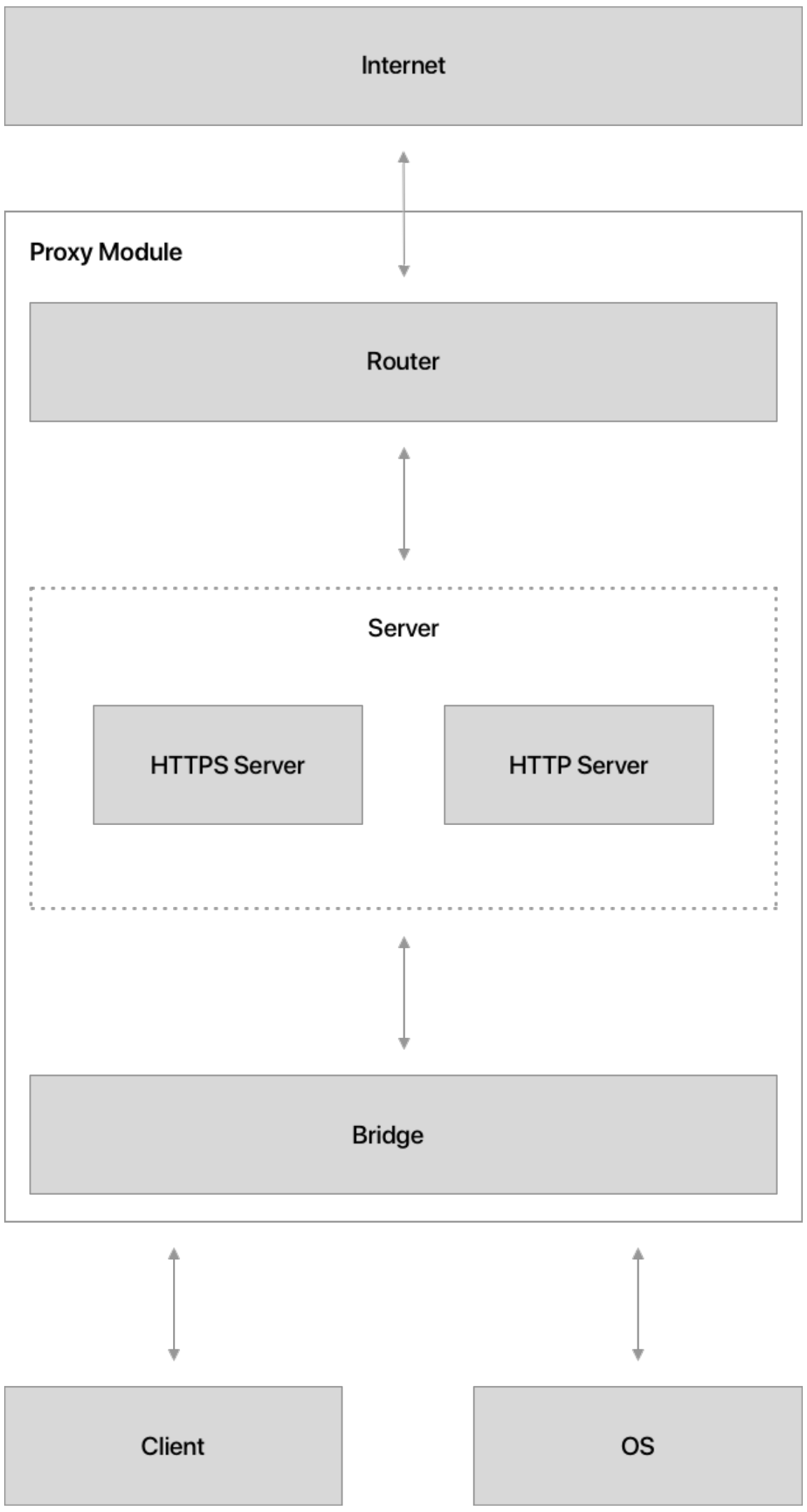

Fig. 4: Proxy Framework Architecture

- Both HTTP and HTTPS servers are almost identical. The only difference is the certificate handling that is required by HTTPS, since the requests are encrypted. Aside from this certificate management, the remaining request handling is identical and it is performed by the Server abstraction, which is inherited by both servers. To be able to decipher HTTPS requests, Robin's proxy module acts as a *Man In the Middle* proxy. The user must trust Robin's certificates to read HTTPS requests. These HTTPS requests are exchanged between the proxy module and the target server using the server certificates and then they are exchanged with the user using Robin's certificates. This way it is possible to easily decipher the package encryption and read/change its content.
- All requests are sent and received by their final destination through the Router module. This is the connection between the framework and the internet.

This framework is fundamental to Robin's development since its purpose is based on network traffic sniffing and manipulation. The Proxy framework is located at the Main module since it requires communication on the OS-level. Once the packages are intercepted they are sent to the Renderer module, where they can be presented and manipulated by the user.

## VI. Actual Case Study

On this section, we apply the tool in a real case scenario. The target website is from a famous news company in Brazil[17]. In this section, we demonstrate how to use Robin to gain full access to random customers accounts, extract sensitive information such as bank accounts, credit card, email, addresses, and much more. At the end, we discuss future attack scenarios using other modules presented in Section IV. The pentest process is divided into two stages:

- *Analyzes* - At this stage Robin is used to analyze the target network requests, looking for possible breaches on the data that is being exchanged. At this point we do not modify the information, we just analyze and point out hypothesis about possible exploit scenarios based on the data being exchanged.
- *Execution* - On this stage, we try the hypotheses created on the previous stage. Then the data being exchanged are modified to see how the target application responds to that.

It is important to emphasize that **all sensitive information**, such as the target website and the hacked user was **anonymized and protected**. The goal of this use case is to demonstrate the tool usage on a real-world scenario, and not to expose sensitive information.

### *A. Analyzes*

The target news company has a news website where customers can purchase plans to read their exclusive newspapers and posts. At this stage, their login system was analyzed. The requests and data exchanged, when the user login on their platform, is recorded by Robin so we can create some hypothesis about any possible breach on their system. Figure 5 shows some data exchanged when the login credentials are sent to their server.

At this point, we have no evidence of a flaw in the login system. The response returns some verification tokens, the customer name, email and other flags. We also point out that their system is using Google Authentication API to perform the customers authentication. This is important because our chances of finding some exploit naturally decrease since they are using a solid and well-known authentication API.

[17]The security engineers of the company have already been notified on these vulnerabilities.

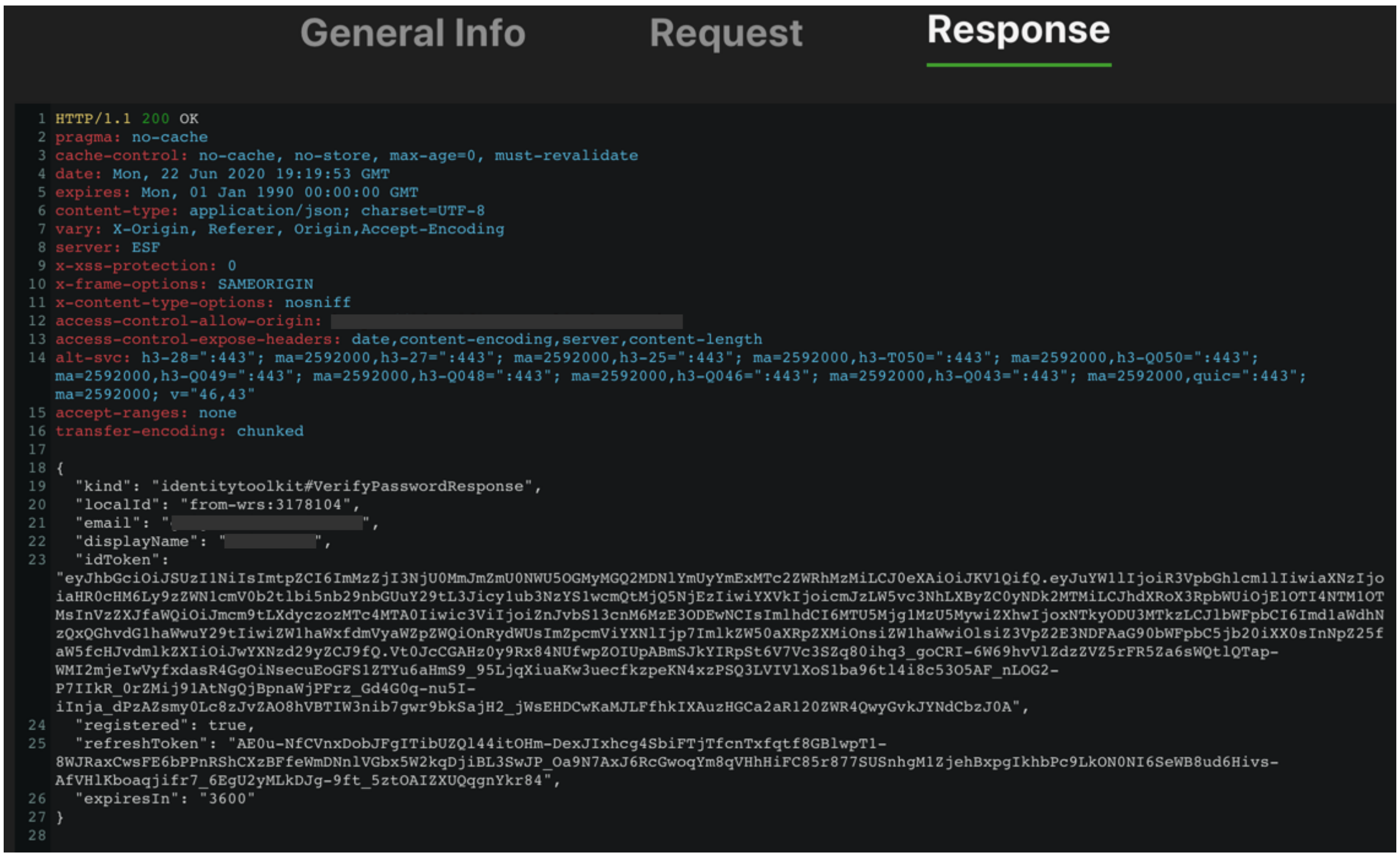

Fig. 5: Login Request

Also, identifiers and tokens are always interesting to take a closer look, since this kind of information is commonly used to perform Insecure Direct Object References (IDOR) [18] exploits.

Once the login information is exchanged, the application starts firing requests to fetch user information and other application data. Figure 6 shows these requests content. Taking a closer look at these requests content, we can notice some coincidence with the previously analyzed request. Both of them use a particular identifier, prefixed with *from-wrs*, followed by a number. Coincidentally, this is the same identifier that was returned by the authentication request previously made.

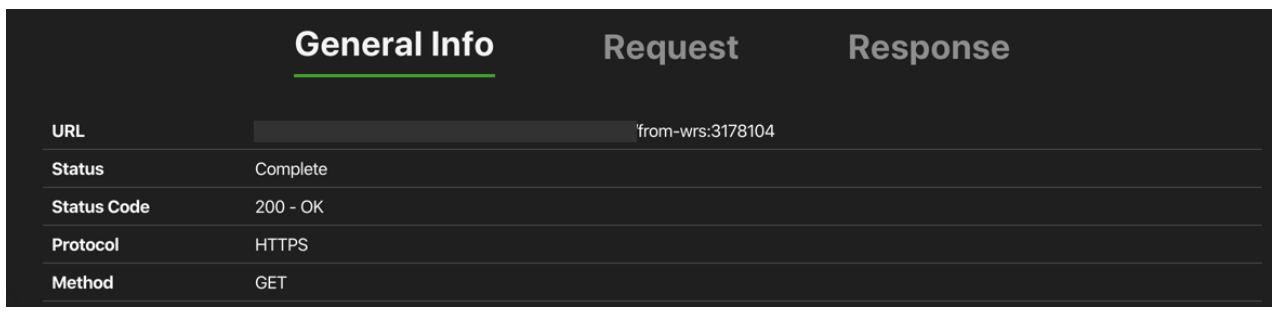

Fig. 6: User Info Request

[18]https://www.acunetix.com/blog/web-security-zone/what-are-insecure-direct-object-references/

This raises some unanswered and really interesting questions:

- Why the requests that fetch the user information are not using Google's API tokens returned from the previous request?
- We know that session tokens must be completely random to guarantee the session to be safe from any possible brute-force attacks, and the id *3178104* does not seem to be random at all. Moreover, this id looks like an incremental identifier, which is even worst.

The most obvious hypothesis - that is later proven right - is that this id returned from Google's API is the user identifier on the magazine server. Although their authentication system uses Google's Auth API, they are not well integrated, so the website is using this user identifier as the "authentication" token to fetch the information instead of the tokens returned from Google.

This is a clear situation of IDOR (Insecure Direct Object Reference) exploit, where the exchanged objects can be manipulated to gain improper access. The next stage tries this knowledge to exploit this possible vulnerability.

## B. Execution

The exploit scenario is clear. Unless something was wrongly analysed in the previous stage, this is an clear situation of an IDOR exploit, and we can change that identifier to get improper access to the system.

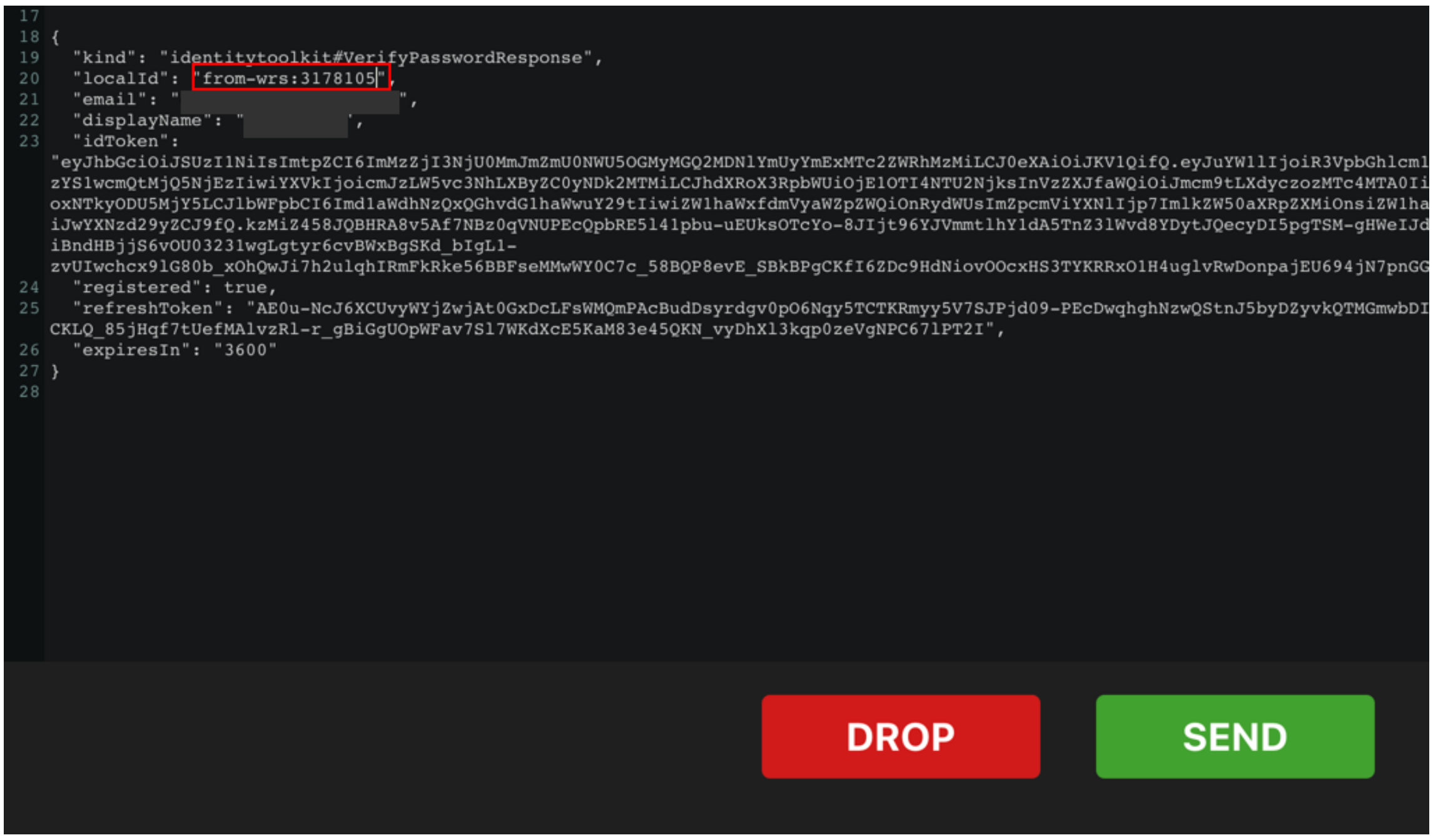

Fig. 7: Response Manipulation

Robin provides features to manipulate the requests before they leave the machine and the responses before they are consumed by the client, and we will use these features to manipulate the identifier. Figure 7 shows the response manipulation from Google's Authentication API. If the hypothesis is correct, this id is an incremental identifier and it will be used by the system to indicate which user is asking for information. Just increasing this identifier by 1 may return the information from a completely random user from the platform.

Tapping the Send button from Robin will make the response arrive modified to the client. This way our browser will consume the identifier *3178105* instead *3178104*. Once the response is consumed, the requests that fetch the user information are sent and, as shown by Figure 8, the requests that fetch the user information are being made using the recently modified identifier *3178105*. It indicates that the application accepted our modified response and that we now are logged in as user *3178105*. Besides that, the response from the request returns data from a customer called *Marlene*, which is not the display name of our test account.

This proves the IDOR exploit on the company website. Just by modifying a simple identifier, we could get full access from a random user account. Navigating a little bit through their settings page, we could extract information like full name, email address, **phone number**, **bank account**, **customer house address**, the days that the physical newspaper is delivered to the customer house, **customer credit card** and much more sensitive information. Such a small development issue allowed us to hack this customer account using Robin to manipulate the requests and response information.

This entire attack scenario was described in a formal document sent to the news company so they can fix it as soon as possible. IDOR exploits are classified as *Injection* attacks and are #1 Applications Security Risk according to OWASP Top 10 project [8].

This case study is just an example on how powerful Robin can be just by having its core proxy module. We have demonstrated just one possible attack scenario, but Robin is also able to perform a lot of different vulnerabilities testing, such as *Cross-Site Scripting (XSS)*, *Cross-Site Request Forgery (CSRF)*, *Unvalidated Redirects and Forwards* and so on. The possibility of being able to read and change HTTP and HTTPS requests gives a lot of possible different attack scenarios, and Robin is ready to handle all of them.

## VII. FUTURE DEVELOPMENT

With the features provided by the current Robin version, we could analyze and execute a simple attack scenario. We discovered an IDOR exploit on the target website and could get unauthorized access to a random customer account, and there, we could extract a lot of sensitive information. By itself this breach on the authentication flow is critical, but we can go beyond that.

As described in Section IV and illustrated by Figure 2, this demonstrates the capability that only the **Proxy**

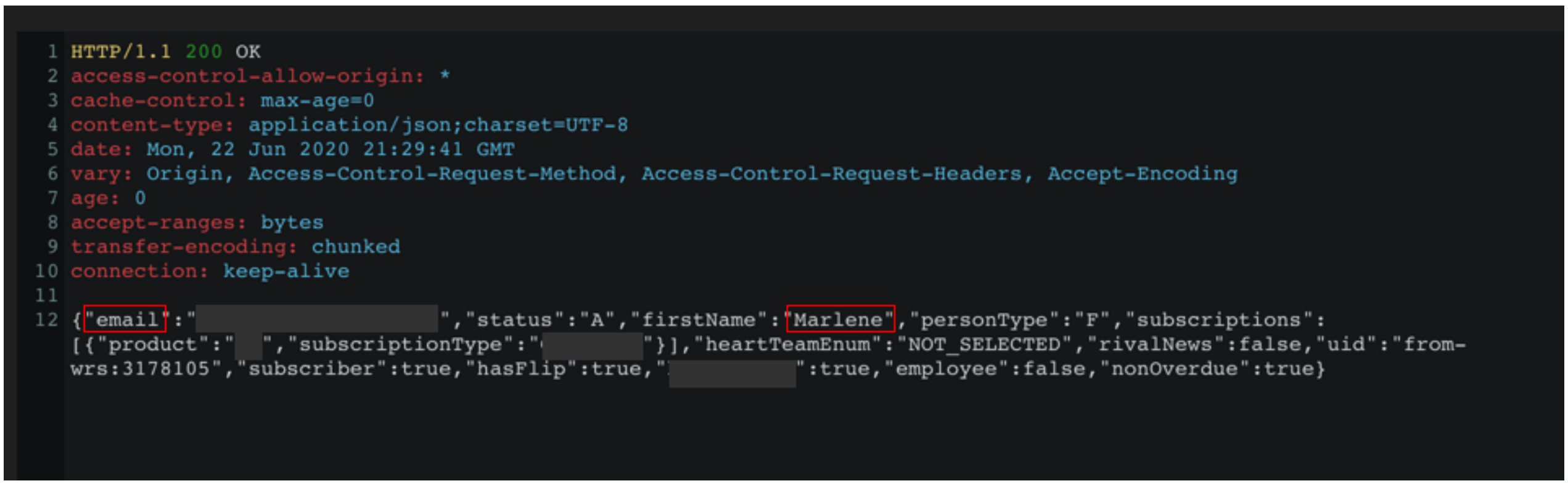

Fig. 8: User Information Response

module allowed us to achieve. But with the information that we extracted with this attack, combined with the **Repeater** module described in the same section, we could create an attack scenario where we performed multiple requests incrementing the user identifier on authentication requests to log in on different accounts and extract these accounts data. Combining both modules with the discovered exploit it could be possible to **extract the information from all platform users, including personal and payment data**. And this attack scenario could be possible without writing a single line of code. Robin can be able to perform all these tasks by itself.

The development of the Proxy module is just the beginning. Combining the 5 presented modules, Robin can become a very powerful tool able to perform complex attack scenarios quickly and precisely. Robin is an open-source project, and its current development stage can be found on its Github page [19]

## VIII. CONCLUSIONS

This paper introduces Robin: a powerful open-source tool that helps information security professionals and developers to find possible vulnerabilities on web applications. The application development environment and architecture were discussed and a real-world exploit scenario was demonstrated using Robin. We aimed at the discussions of the previous sections to emphasize and to demonstrate the importance of information security research and knowledge when developing web applications. Minor security issues may lead to several consequences.

We also discussed the information security tools available on the market and their pros and cons. Robin was created aiming to put together the best modules available on these tools, remove or fix the bad ones and to be a completely free and open-source tool, driven by the information security community. We hope that with this contribution we can help to attract more developers to the security research world and make the applications safer to its users.

[19]http://github.com/olmps/robin

Some future work could be to add some formal specification of attacks that could be interpreted by Robin in order to verify whether a tool provides the security as it was specified, similar to the one presented by Peralta et al. [22]

**Acknowledgements**. This is a report for the Final Project for the Computer Science Course from the Pontifical Catholic University of Rio Grande do Sul, Brazil. A revised version of this paper was submitted to a conference. We thank the reviewers of the project, Prof. Cesar De Rose and Prof. Ana Benso, for their comments. Avelino F. Zorzo is supported by CNPq/Brazil.